\documentclass[aps, prl, reprint, showpacs, twocolumn, 10pt, groupedaddress]{revtex4-1}

\usepackage{graphicx, amsmath, dsfont, dcolumn, bm, times, color, braket, amssymb, multirow}
\usepackage{ulem}

\begin{document}

\bibliographystyle{apsrev4-2}

\title{Meissner effect in non-Hermitian superconductors}

\author{Shun Tamura, Helene M\"uller, Linus Aliani, and Viktoriia Kornich}
\affiliation{Institute for Theoretical Physics and Astrophysics, University of W\"urzburg, 97074 W\"urzburg, Germany }

\date{\today}

\begin{abstract}
We study theoretically Meissner effect in non-Hermitian systems of BCS type, i.e., with an electron-electron interaction leading to the mean field description in a Cooper channel, via superfluid stiffness. We show that depending on the values of the mean fields, chemical potential, and temperature, we obtain paramagnetic or diamagnetic Meissner effect. Notably, positive real part of the product of mean fields guarantees diamagnetic Meissner effect in an $s$-wave 3D non-Hermitian superconductor. Once the mean fields are close to being anti-Hermitian, 2D $s$-, $p_x$-, and $d$-wave superconductors exhibit interesting behaviour, including negative superfluid stiffness giving paramagnetic Meissner effect for certain parameters. Negative superfluid stiffness indicates instability of this collective state.
\end{abstract}

\maketitle

\let\oldvec\vec
\renewcommand{\vec}[1]{\ensuremath{\boldsymbol{#1}}}
{\it Introduction.--} Non-Hermitian systems are currently of high interest due to new effects, e.g., non-Hermitian skin effect \cite{alvarez:prb18, yao:prl18, torres:jpm20, zhang:natcom22, kawabata:prx23}, non-Hermitian topological effects \cite{okuma:anrev23}, effects related to exceptional points \cite{bergholtz:rmp21}, new type of mathematical models \cite{mostafazadeh:jmp02}, and other interesting results \cite{wu:AAPPS22, liu:jpcm21, yerin:prb23, kruchkov:arXiv23} they allow for. There are experimental results confirming some of the theoretical models \cite{zhang:natcom21, liu:prr23}, but the mathematical formalism of non-Hermitian quantum mechanics is still under debate \cite{kornich:prl23, meden:rop23}. Non-Hermiticity is often a result of a system being out of equilibrium or, in other words, under external influence. 

Superconductivity is a striking phenomenon that is still not understood because its physical grounds in unconventional superconductors are usually not known \cite{pelc:sciadv19, gingras:prb22}. Enhancement of superconductivity or its induction is mainly studied with light pumping, but also with strain and acoustic waves \cite{fausti:science11, mitrano:nature16, seo:asiamater20, kornich:scipost22}. These external effects can be theoretically taken into account in terms of non-equilibrium techniques, such as Keldysh Green's functions \cite{kornich:scipost22}, but can also be considered in a visibly more simple (but also less precise) way, i.e., in terms of non-Hermitian technique. 

In this work, we study non-Hermitian Hamiltonians of BCS type, i.e., mean-field representation of non-Hermitian superconductors, where the mean fields are not Hermitian conjugate.  Our goal is to study, what kind of Meissner effect these systems have. In equilibrium, diamagnetic Meissner effect (magnetic field expulsion) is considered to be one of the criteria for a bulk material to be a stable superconductor. Yet, paramagnetic Meissner effect is suspected to occur due to the formation of odd-frequency superconductivity in bulk \cite{heid:physb95, solenov:prb09} and in junctions \cite{yokoyama:prl11, mironov:prl12, dibernardo:prx15, tanaka:prb15, lee:prb17}. Therefore, the definition of a superconductor is somewhat ambiguous in this sense.  

In order to study Meissner effect, we consider superfluid stiffness, $Q$, that denotes the rigidity of the state described by the given BCS-type Hamiltonian against fluctuations of the phase of the condensate of Cooper pairs. We assume, there is a condensate of Cooper pairs, if $\langle \psi\psi\rangle\neq 0$ and $\langle\psi^\dagger\psi^\dagger\rangle\neq 0$, where $\psi$ and $\psi^\dagger$ are annihilation and creation operators of electrons in the system, respectively. $Q$ is related to the London penetration length as $Q\propto \lambda^{-2}$. Thus, if $Q<0$, the penetration length becomes imaginary, and magnetic field oscillates within the material instead of exponentially decaying. This means, the Meissner effect is not diamagnetic. We will call it paramagnetic Meissner effect \cite{schmidt:prb20}. 

In a non-Hermitian system, $Q$ can have both real and imaginary parts~\cite{xian:scipost24}. In our model, the imaginary part of $Q$ is related to the decay and/or pumping of Bogoliubov quasiparticles. The real part denotes rigidity of a phase of a condensate, i.e., if it is energetically favourable to have all particles in phase or not. When not, i.e. $\Re[Q]<0$, we obtain paramagnetic Meissner effect. We have found that $\Re[\bar{\Delta}\Delta]>0$, where $\bar{\Delta}$ and $\Delta$ are the s-wave mean fields, is a strong criterium for the diamagnetic Meissner effect in 3D. If not, depending on the chemical potential, we can obtain paramagnetic Meissner effect. We also consider numerically $s$-, $p_x$-, and $d$-wave superconductors on a square lattice. We show that $\Delta$ close to $-\bar{\Delta}$ is a particular region for $Q$.    

{\it Derivation of mean-field representation.--} We start from considering a system consisting of electrons and some kind of bosons, e.g., phonons, photons or excitons, that interact with electrons and are non-equilibrium in general. Then, the partition function is:
\begin{eqnarray}
\mathcal{Z}=\int \mathcal{D}[\bar{\psi},\psi,\bar{\varphi},\varphi]e^{-[S_{\rm el}(\bar{\psi},\psi)+S_{\rm int}(\bar{\psi},\psi,\bar{\varphi},\varphi)+S_{\rm bos}(\bar{\varphi},\varphi)]},\ \ \ \ \ \ 
\end{eqnarray} 
where $\varphi$ and $\bar{\varphi}$ are the bosonic fields, $S_{\rm el}$, $S_{\rm bos}$, and $S_{\rm int}$ are the parts of action describing the system of electrons, bosons, and their interaction, respectively. Once, we integrate out bosonic degrees of freedom, $\bar{\varphi}$ and $\varphi$, we obtain $S_{\rm e-e}$ that is an electron-electron interaction mediated by the non-equilibrium bosons:
\begin{eqnarray}
&&S_{\rm e-e}=\ \ \ \ \ \\ \nonumber&&-\sum_{\sigma,\sigma'}\int V_{\tau-\tau',{\bf r}-{\bf r'}}\bar{\psi}_{\tau,{\bf r},\sigma}\bar{\psi}_{\tau',{\bf r'},\sigma'}\psi_{\tau',{\bf r'},\sigma'}\psi_{\tau,{\bf r},\sigma}d{\bf r}d{\bf r'}d\tau d\tau',
\end{eqnarray}
where indices $\tau$ and $\tau'$ denote time in imaginary representation, ${\bf r}$ and ${\bf r'}$ are spatial coordinates, and $\sigma$, $\sigma'$ denote spin projections $\uparrow$ and $\downarrow$. The electron-electron interaction potential $V$ contains bosonic Green's functions and therefore it is usually complex for non-equilibrium bosons \cite{SM}. Thus, $V$ acquires unconventional properties, if bosons have unusual characteristics, e.g., $V$ can be odd in momentum \cite{kornich:prr22}. However we do not specify the exact model for bosons here, because complex $V$ leads to non-Hermitian mean fields in general.

In order to obtain the mean field representation of the action, $S_{\rm eff}$, we perform Hubbard-Stratonovich transformation in a Cooper pair channel \cite{SM}:
\begin{eqnarray}
\label{eq:DeltaHS}
\Delta_{{\bf k},{\bf q}}^{\sigma',\sigma}&\rightarrow& \Delta_{{\bf k},{\bf q}}^{\sigma',\sigma}+\frac{1}{\sqrt{\beta L^d}}\sum_{{\bf k}'}V_{{\bf k}-{\bf k}'}\psi_{-{\bf k}'+{\bf q},\sigma'}\psi_{{\bf k}',\sigma},\\
\label{eq:barDeltaHS}
\bar{\Delta}_{{\bf q},{\bf k}}^{\sigma,\sigma'}&\rightarrow& \bar{\Delta}_{{\bf q},{\bf k}}^{\sigma,\sigma'}+\frac{1}{\sqrt{\beta L^d}}\sum_{{\bf k}'}\bar{\psi}_{{\bf k'},\sigma}\bar{\psi}_{-{\bf k}'+{\bf q},\sigma'}V_{{\bf k}'-{\bf k}},
\end{eqnarray}
where indices ${\bf k}$, ${\bf q}$ denote 4-vectors and comprise Matsubara frequency and momentum, $\beta$ is the inverse temperature, $L$ is the size of the system with dimension $d$. From Eqs.~(\ref{eq:DeltaHS}) and (\ref{eq:barDeltaHS}) follows that $\bar{\Delta}$ and $\Delta$ are not Hermitian conjugate in general. In the following, we restrict our consideration to zero-center-of mass momentum of Cooper pairs. We note, that this Hubbard-Stratonovich transformation is valid only if the Hermitian part of $V^{-1}$ is positive definite \cite{SM}. If not, we suggest to use $\langle AB\rangle\simeq A\langle B\rangle+\langle A\rangle B-\langle A\rangle\langle B\rangle$, although it is not exact in contrast to Hubbard-Stratonovich transformation.

In time and coordinate representation, $S_{\rm eff}$ is
\begin{eqnarray}
\nonumber&&S_{\rm eff}=\\ \nonumber
&&\sum_{\sigma}\int d{\bf x}\left\{\left[\bar{\psi}_{{\bf x},\sigma}\left(\partial_\tau+\frac{(-i\boldsymbol{\nabla}-e{\bf A})^2}{2m}-\mu\right)\psi_{{\bf x},\sigma}\right]+\right.\\ \nonumber&&\left.\sum_{\sigma'}\left(\int d{\bf x}'\left[\bar{\Delta}^{\sigma,\sigma'}_{{\bf x}-{\bf x}'}\psi_{{\bf x}',\sigma'}\psi_{{\bf x},\sigma}+\bar{\psi}_{{\bf x},\sigma}\bar{\psi}_{{\bf x}',\sigma'}\Delta^{\sigma',\sigma}_{{\bf x}-{\bf x}'}\right]+\right.\right.\\ \label{eq:Seff}&&\left.\left.\beta L^d\bar{\Delta}^{\sigma,\sigma'}_{{\bf x}}V^{-1}_{{\bf x}}\Delta_{{\bf x}}^{\sigma',\sigma}\right)\right\},
\end{eqnarray}
where ${\bf x}=(\tau,{\bf r})$. Here, the mean fields $\bar{\Delta}_{\tau,{\bf r}}=|\bar{\Delta}_{\tau,{\bf r}}|e^{i\bar{\phi}_{\tau,{\bf r}}}$ and $\Delta_{\tau,{\bf r}}=|\Delta_{\tau,{\bf r}}|e^{i\phi_{\tau,{\bf r}}}$ are not necessarily Hermitian conjugate, $m$ is the effective mass of electron, $\bf{A}$ is a vector potential, $\mu$ is the chemical potential.

{\it Superfluid stiffness in non-Hermitian case.--} Even though non-Hermiticity is often a result of the absence of equilibrium, non-Hermitian systems should still be gauge-invariant. If we perform gauge transformation
$\psi_{{\bf r}}\rightarrow \psi_{{\bf r}}e^{i\alpha({\bf r})}$ and
${\bf A}\rightarrow {\bf A}+\frac{\boldsymbol{\nabla} \alpha({\bf r})}{e}$
we need to also do 
$\Delta_{{\bf r}-{\bf r}'}\rightarrow e^{i(\alpha({\bf r})+\alpha({\bf r}'))}\Delta_{{\bf r}-{\bf r}'}$ and
$\bar{\Delta}_{{\bf r}-{\bf r}'}\rightarrow e^{-i(\alpha({\bf r})+\alpha({\bf r}'))}\bar{\Delta}_{{\bf r}-{\bf r}'}$.
The gauge-invariant term that denotes the energy cost for deforming the phase is
\begin{eqnarray}
\nonumber S_\phi=Q\int \frac{d\tau d\tau'd{\bf r}d{\bf r}'}{16e^2\beta L^d}\{(\boldsymbol{\nabla}_{\bf R}\bar{\phi}_{{\bf r},{\bf r}'}+e[{\bf A}({\bf r})+{\bf A}({\bf r}')])^2+\\ +(\boldsymbol{\nabla}_{\bf R}\phi_{{\bf r},{\bf r}'}-e[{\bf A}({\bf r})+{\bf A}({\bf r}')])^2\},\ \ \ \ \ 
\end{eqnarray}
where ${\bf R}=({\bf r}+{\bf r}')/2$ and ${\bf r}-{\bf r}'$ form the other basis. If we consider the uniform twist of the phase, $\alpha({\bf r})={\bf a}\cdot {\bf r}$, we can remove it by the transformation
\begin{eqnarray}
\Delta_{{\bf r},{\bf r}'}&\rightarrow& e^{-i2{\bf a}\cdot{\bf R}}\Delta_{{\bf r},{\bf r}'}=\Delta_{{\bf r}-{\bf r}'},\\
\bar{\Delta}_{{\bf r},{\bf r}'}&\rightarrow& e^{i2{\bf a}\cdot{\bf R}}\bar{\Delta}_{{\bf r},{\bf r}'}=\bar{\Delta}_{{\bf r}-{\bf r}'},\\
{\bf A}&\rightarrow& {\bf A}-\frac{{\bf a}}{e}.
\end{eqnarray}
Thus, the phase twist is equivalent to the shift of vector-potential, and using just a shift of vector-potential we can find $Q$.

{\it Calculation of superfluid stiffness.--} In this section, we will consider constant $\Delta$ and $\bar{\Delta}$ for simplicity. If we integrate out fermionic degrees of freedom $\bar{\psi}$ and $\psi$ from the partition function, we obtain in momentum representation
\begin{eqnarray}
\nonumber
S=-\sum_{{\bf k},n}{\rm Tr}\ln{\begin{pmatrix}i\omega_n-\frac{({\bf k}-e{\bf A})^2}{2m}+\mu & -\Delta\\ -\bar{\Delta} & i\omega_n+\frac{({\bf k}+e{\bf A})^2}{2m}-\mu\end{pmatrix}}\\ 
\label{eq:S} +\sqrt{\beta L^d}\bar{\Delta}\Delta V^{-1}.\ \ \ \ \ \ \ 
\end{eqnarray}
Analogously to the calculation in Ref.~\cite{coleman:book}, but for $S$ from Eq.~(\ref{eq:S}) with non-Hermitian mean fields, we obtain
\begin{eqnarray}
\nonumber Q^{\alpha\beta}&=&\frac{T}{L^3}\frac{\partial^2 S}{\partial A_\alpha\partial A_\beta}\Big|_{{\bf A}=0}=\\ &=&\frac{4e^2T}{L^3}\sum_{{\bf k}, n}\frac{\bar{\Delta}\Delta\nabla_\alpha \varepsilon_{\bf k}\nabla_\beta\varepsilon_{\bf k}}{(\varepsilon_{\bf k}^2+\bar{\Delta}\Delta+\omega_n^2)^2},\ \ \ \ \
\end{eqnarray}
where $\varepsilon_{\bf k}=k^2/(2m)-\mu$. 

\begin{figure}[tb]
	\begin{center}
		\includegraphics[width=0.8\linewidth]{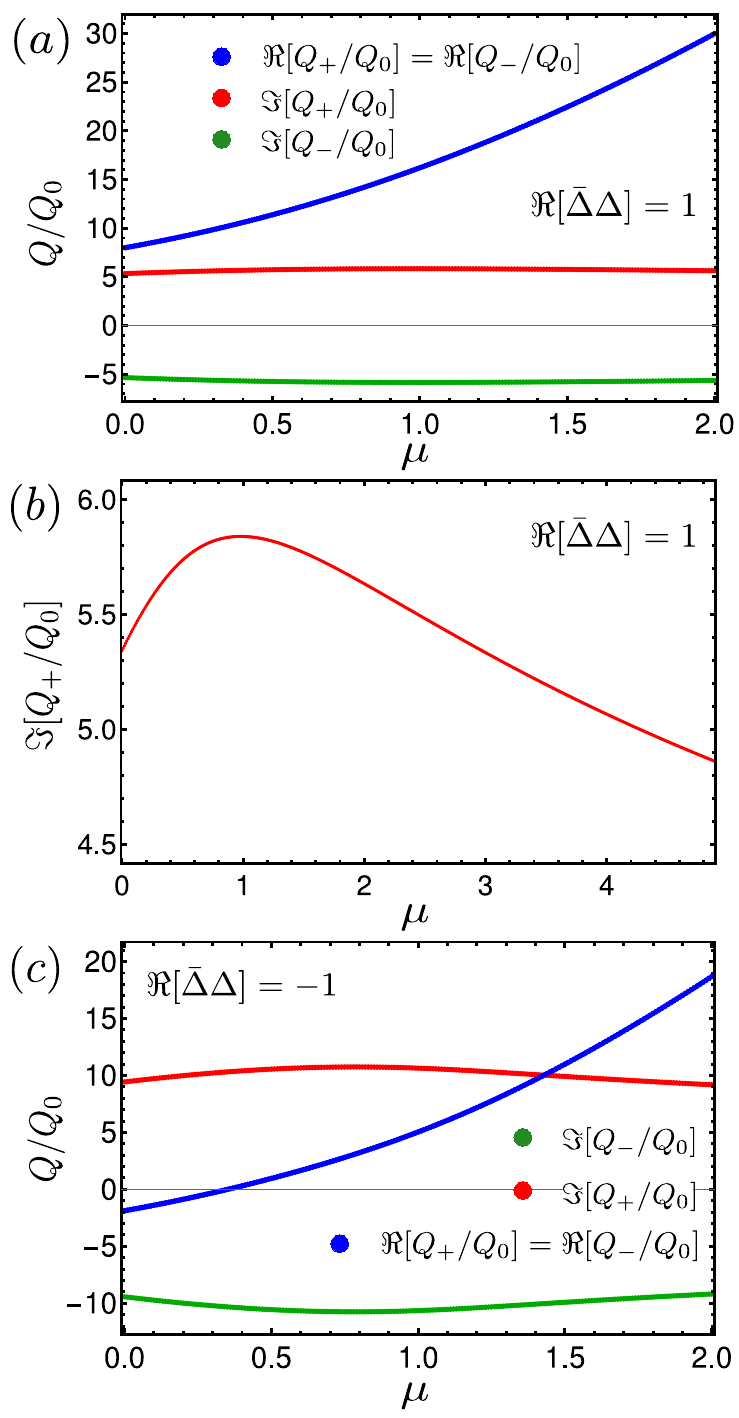}
		\caption{Real and imaginary parts of superfluid stiffness $Q$ normalized by $Q_0=e^2 \sqrt{2m}/(6\pi^2)$. The indices $\pm$ in $Q_{\pm}$ denote $\bar{\Delta}\Delta$ with positive and negative imaginary parts, respectively. (a) Real and imaginary parts of $Q_\pm/Q_0$ at $\bar{\Delta}\Delta=1\pm i$. We show that $\Re[Q_+]=\Re[Q_-]$ and $\Im[Q_+]=-\Im[Q_-]$. The real parts grow with $\mu$, because the larger number of charge carriers provide stronger stiffness of the phase. (b) Imaginary part of $\Im[Q_+/Q_0]$ has maximum at $\mu\neq 0$, here $\bar{\Delta}\Delta=1\pm i$. (c) Real and imaginary parts of $Q_\pm/Q_0$ at $\bar{\Delta}\Delta=-1\pm i$. Again, $\Re[Q_+]=\Re[Q_-]$ and $\Im[Q_+]=-\Im[Q_-]$, but the real part has an interval of $\mu$, where it is negative, indicating paramagnetic Meissner effect.}
		\label{fig:ReImQ}
	\end{center}
\end{figure}

Once we transfer to spherical coordinates, we see that $Q^{\alpha\beta}=0$ for $\alpha\neq\beta$. As we have no anisotropy, we denote $Q^{\alpha\alpha}$ as simply $Q$. We first sum over momenta analytically, and then numerically integrate over Matsubara frequency, taking into account that $T\sum_n\rightarrow \int d\omega/(2\pi)$ at zero temperature. 

We have plotted real and imaginary parts of $Q/Q_0$, where $Q_0=e^2\sqrt{2m}/(6\pi^2)$ in Fig.~\ref{fig:ReImQ}. We consider $\bar{\Delta}\Delta=1\pm i$ in Figs.~\ref{fig:ReImQ} (a) and (b), and $\bar{\Delta}\Delta=-1\pm i$ in Fig.~\ref{fig:ReImQ} (c). We denote $Q_\pm$ the stiffness with positive and negative $\Im[\bar{\Delta}\Delta]$, respectively. We can see that in Fig.~\ref{fig:ReImQ}, $\Im[Q_+]>0$ and $\Im[Q_-]<0$ and $\Im[Q_+]=-\Im[Q_-]$. The imaginary part of $Q$ comes from decay or pumping of Bogoliubov quasiparticles. This can be seen from Eq.~(\ref{eq:Seff}), if we diagonalize it in fermionic degrees of freedom in momentum space, i.e. use Bogoliubov quasiparticle basis. The eigenvalues are then $E_{1,2}=\pm \sqrt{(\frac{k^2}{2m}-\mu)^2+\bar{\Delta}\Delta}$. Only the second term under square root can be imaginary or negative. Thus, the sign of the imaginary part of $E_{1,2}$ is defined by ${\rm sign}(\Im[\bar{\Delta}\Delta])$, where ${\rm sign}[x]$ is $1$ for $x\geq 0$ and $-1$ for $x<0$.

For $\bar{\Delta}\Delta=1\pm i$, we obtain that $\Re[Q_+]=\Re[Q_-]>0$ and grow further with $\mu$. The growth of the real part corresponds to the increase of the number of charge carriers involved in the condensate, when $\mu$ grows \cite{coleman:book}.

However, it can also be that $\Re[Q]\leq 0$, see e.g., Fig.~\ref{fig:ReImQ} (c), where $\bar{\Delta}\Delta=-1\pm i$. This means that for some parameters, the twist of phase is energetically favourable, consequently, it is energetically favourable to {\it not} have all the particles in phase. In other words, any fluctuation of phase would be amplified and there is no stable superconductivity in the sample. In London limit, this means that the penetration length of a magnetic field is imaginary, meaning, the magnetic field penetrates through the material and oscillates there, and such Meissner effect we call paramagnetic.

Let's consider the signs of $\Re[Q]$ and $\Im[Q]$ in more detail (for the full derivation, see Supplemental Material~\cite{SM}). As we see from Fig.~\ref{fig:ReImQ}, $\Re{[Q]}$ grows with $\mu$. Therefore, we consider $\mu=0$ in order to define, when $\Re[Q]$ has a change of sign.
\begin{eqnarray}
\nonumber Q(\mu=0)=\frac{ie^2\sqrt{2m}\bar{\Delta}\Delta}{6\pi^2}{\rm sign}^*[\Im[\bar{\Delta}\Delta]]\times \\ \int d\omega \left[\frac{(-\bar{\Delta}\Delta-\omega^2)^{1/4}-\sqrt{-\sqrt{-\bar{\Delta}\Delta-\omega^2}}}{\bar{\Delta}\Delta+\omega^2}\right],
\end{eqnarray}
where the function ${\rm sign}^*[x]$ is $1$ for $x>0$ and $-1$ for $x\leq 0$. Once we expand all the square roots, we obtain that at $\Re[\bar{\Delta}\Delta]>0$, $\Re[Q(\mu=0)]>0$. If $\Re[\bar{\Delta}\Delta]\leq 0$, there is an interplay of terms under integral over $\omega$ and the outcome is not known in general. Thus, if $\Re[\bar{\Delta}\Delta]>0$, we obtain only positive $\Re[Q]$.

Similarly for $\Im[Q(\mu=0)]$, if $\Re[\bar{\Delta}\Delta]>0$, we obtain ${\rm sign}(\Im[Q(\mu=0)])={\rm sign}^*(\Im[\bar{\Delta}\Delta])$. Analogously as for the real part, if $\Re[\bar{\Delta}\Delta]\leq 0$, the sign of $\Im[Q(\mu=0)]$ is not known in general due to interplay of different terms under integral over $\omega$.

  \begin{figure}[t]
    \begin{center}
      \includegraphics[width=8.6cm]{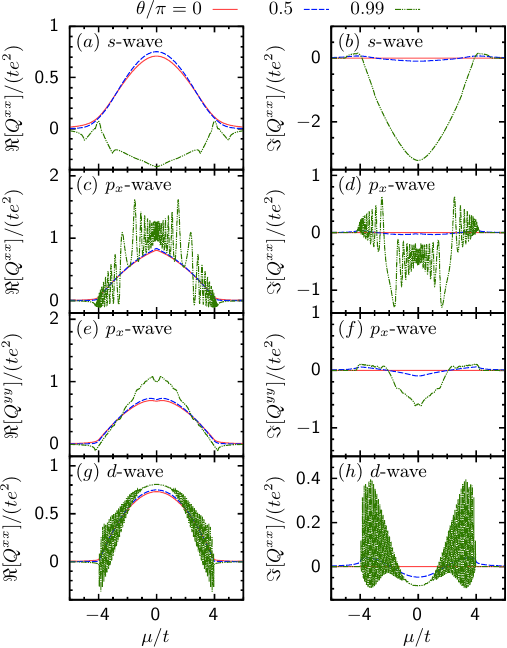}
    \end{center}
    \caption{
      $Q$ is plotted as a function of $\mu$ for (a,b) $s$-wave, (c,d,e,f) $p_x$-wave, and (g,h) $d$-wave superconductors with $\theta/\pi=0$, $0.5$, and $0.99$.
      The real parts are shown in (a,c,e,g), and the imaginary parts are shown in (b,d,f,h).
      $Q^{xx}$ is shown in (a,b,c,d,g,h), and $Q^{yy}$ is shown in (e,f).
      $\beta t=10^2$ and $\Delta_0/t=1$ for all plots.
  }\label{fig:mu_dep_spd}
  \end{figure}
  \begin{figure}[t]
    \begin{center}
      \includegraphics[width=8.6cm]{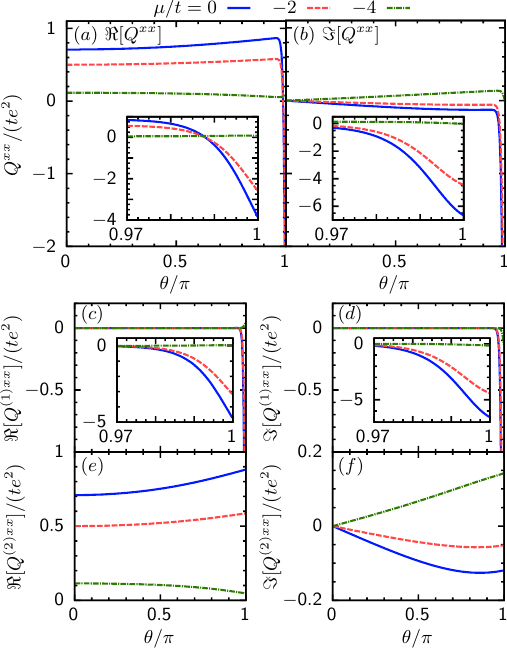}
    \end{center}
    \caption{
      (a) $\Re[Q^{xx}]$, (b) $\Im[Q^{xx}]$, (c) $\Re[Q^{(1)xx}]$, (d) $\Im[Q^{(1)xx}]$, (e) $\Re[Q^{(2)xx}]$, and (f) $\Im[Q^{(2)xx}]$ is plotted as a function of $\theta$ for $\mu/t=0$, $-2$, and $-4$ with $\Delta_0=t$ and $\beta t=10^2$ for $s$-wave superconductor.
      The insets in (a), (b), (c), and (d) show the magnified views close to $\theta=\pi$.
  }\label{fig:theta_dep}
  \end{figure}
  \begin{figure}[t]
    \begin{center}
      \includegraphics[width=8.6cm]{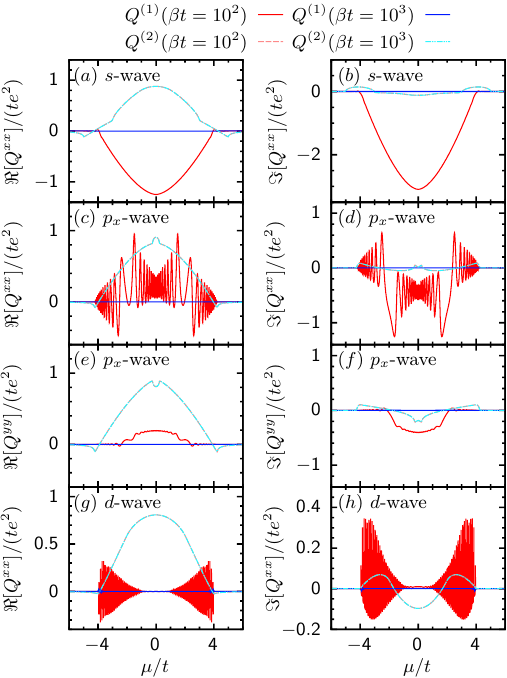}
    \end{center}
    \caption{
      $Q^{(1)}$ and $Q^{(2)}$ at $\beta t=10^2$ and $10^3$ are plotted as a function of $\mu$ with $\theta/\pi=0.99$ and $\Delta_0/t=1$.
      (a,b) $s$-wave, (c,d,e,f) $p_x$-wave, and (g,h) $d$-wave superconductors.
      The real parts are shown in (a,c,e,g), and the imaginary parts are shown in (b,d,f,h).
      $Q^{(1,2)xx}$ is shown in (a,b,c,d,g,h), and $Q^{(1,2)yy}$ is shown in (e,f).
  }\label{fig:mu_dep_spd_temp}
  \end{figure}
  {\it Stiffness of s-, p-, and d-wave non-Hermitian superconductors on a square lattice.--}
  Let us consider the BCS-type Hamiltonian on the square lattice given by 
  \begin{eqnarray}
  H_\mathbf{k}(\mathbf{A})&=&\begin{pmatrix}\varepsilon_{\mathbf{k}-e\mathbf{A}}&\Delta(\mathbf{k})\\\Delta(\mathbf{k}) e^{i\theta}&-\varepsilon_{\mathbf{k}+e\mathbf{A}}\end{pmatrix},\\
  \varepsilon_{\mathbf{k}\pm e\mathbf{A}}&=&-2t[\cos (k_x\pm eA_x)+\cos (k_y\pm eA_y)]-\mu,\ \ \ \ \ \ \ \ 
  \end{eqnarray}
   with the hopping integral $t$, $0\le\theta<\pi$, and we set the lattice constant as unit of the length.
     The eigenvalues of $H_{\mathbf{k}}(\mathbf{A})$ and the action $S(\mathbf{A})$ are given by
     \begin{align}
       E_{\mathbf{k},\pm}(\mathbf{A})
       =&
       \pm\sqrt{[2t\sum_{j=x,y}\cos k_j\cos(eA_j)+\mu]^2+\Delta^2(\mathbf{k})e^{i\theta}}
       \nonumber\\
        &+2t\sum_{j=x,y}\sin k_j\sin(eA_j),
     \end{align}
     and $S(\mathbf{A})=-\sum_{n,\mathbf{k},\alpha=\pm}\mathrm{Tr}\ln[E_{\mathbf{k},\alpha}(\mathbf{A})-i\omega_n]$, respectively.
     At $\mathbf{A}=0$, $E_{\mathbf{k},\pm}(0)$ can be purely imaginary, and the action can be ill-defined.
     Hence, we avoid discussing at $\theta=\pi$.

  We consider $s$-, $p_x$-, and $d$-wave symmetries of the mean fields. Namely, we take $\Delta(\mathbf{k})$ as: $\Delta(\mathbf{k})=\Delta_0$ for $s$-wave, $\Delta(\mathbf{k})=\Delta_0\sin k_x$ for $p_x$-wave, and $\Delta(\mathbf{k})=\Delta_0(\cos k_x-\cos k_y)$ for $d$-wave cases.
  
  We can decompose $Q^{ab}$ as $Q^{ab}=Q^{(1)ab}+Q^{(2)ab}$, where
  \begin{eqnarray}
  Q^{(1)ab}&=&\frac{1}{L^2}\sum_{{\bf k}, \alpha=\pm}\frac{-\beta}{4\cosh^2{\frac{\beta E_{{\bf k}, \alpha}}{2}}}\frac{\partial E_{{\bf k},\alpha}}{\partial A_a}\frac{\partial E_{{\bf k},\alpha}}{\partial A_b}\Big|_{{\bf A}=0},
  \label{eq:Q1}
  \ \ \ \ \\
  Q^{(2)ab}&=&\frac{1}{L^2}\sum_{{\bf k},\alpha=\pm}\frac{1}{1+e^{\beta E_{{\bf k},\alpha}}}\frac{\partial^2 E_{{\bf k},\alpha}}{\partial A_a\partial A_b}\Big|_{{\bf A}=0}.
  \label{eq:Q2}
  \end{eqnarray}
 
 We find that $Q^{xy}=0$ for all the cases, and $Q^{xx}=Q^{yy}$ for $s$- and $d$-wave superconductors.
  We show the stiffness $Q$ in Fig.~\ref{fig:mu_dep_spd} as a function of $\mu$ with $\theta/\pi=0$, $0.5$, and $0.99$.
    In all cases, $\Im[Q^{xx,yy}]$ is zero at $\theta=0$, and $Q^{xx,yy}$ for $\theta/\pi=0$ and $0.5$ are almost the same.
 For the $s$-wave case, at $\theta/\pi=0.99$, $\Re[Q^{xx}]$ becomes negative in the wide range of $\mu$.
  
  The $\theta$ dependence of $Q$ for the $s$-wave superconductor is shown in Fig.~\ref{fig:theta_dep}.
  $Q^{xx}$ is almost constant with respect to $\theta$ except close to $\theta=\pi$, see Figs.~\ref{fig:theta_dep} (a) and (b). It drastically changes its value close to $\theta=\pi$ for $\mu/t=0$ and $-2$.
  In addition, close to $\theta=\pi$, $\Re[Q^{xx}]$ approaches a negative value for $\mu/t=0$ and $-2$.
    In Figs.~\ref{fig:theta_dep} (c), (d), (e), and (f), $Q^{(1)xx}$ and $Q^{(2)xx}$ are shown.
        From Eq.~\eqref{eq:Q1}, in the low temperature limit, $\beta\rightarrow\infty$, $Q^{(1)xx}=0$ if the energy gap opens, i.e., $|E_{\mathbf{k},\pm}(0)|>0$ for any $\mathbf{k}$.
    For the $s$-wave, the energy gap only closes at $\theta=\pi$, and $Q^{(1)xx}$ at $\beta t=10^2$ is almost zero for $\theta/\pi\lesssim0.97$, but close to $\theta=\pi$, $Q^{(1)xx}$ changes its value [Figs.~\ref{fig:theta_dep} (c) and (d)].
    On the other hand, $Q^{(2)xx}$ shown in Figs.~\ref{fig:theta_dep} (e) and (f), does not drastically change its value close to $\theta=\pi$.
    Hence, negative $\Re[Q^{xx}]$ is realized by $\Re[Q^{(1)xx}]$.
    In Figs.~\ref{fig:mu_dep_spd_temp} (a) and (b), $Q^{xx}$ at $\theta/\pi=0.99$ for $\beta t=10^2$ and $10^3$ is shown.
    If we fix $\theta$, $Q^{(1)xx}$ approaches zero as $\beta t$ increases as explained above.
    $Q^{(2)xx}$ is almost independent on $\beta$, and this can be understood as follows. 
    When $|E_{\mathbf{k},\pm}|>0$ for any $\mathbf{k}$, i.e., energy gap opens, from Eq.~\eqref{eq:Q2}, $1/(1+e^{\beta E_{\mathbf{k},\pm}})\rightarrow 1$ for $\Re[E_{\mathbf{k},\pm}]<0$ and $1/(1+e^{\beta E_{\mathbf{k},\pm}})\rightarrow 0$ for $\Re[E_{\mathbf{k},\pm}]>0$ in the low temperature limit, and $Q^{(2)xx}$ is independent on $\beta$.
    At $\theta/\pi=0.99$, $\beta t=10^2$ is low enough temperature, and $Q^{(2)xx}$ at $\beta t=10^2$ and $10^3$ are almost the same.

    For $p_x$- and $d$-wave superconductors, $Q^{xx}$ shows oscillatory behavior at $\theta/\pi=0.99$ as shown in Figs.~\ref{fig:mu_dep_spd} (c), (d), (g) and (h), and $Q^{xx}$ can have a negative value close to $|\mu|=4t$.
    This oscillatory behavior is suppressed when $\beta$ becomes larger [Figs.~\ref{fig:mu_dep_spd_temp} (c), (d), (g), and (h)].

  For the $p_x$-wave superconductor, $Q^{xx}$ and $Q^{yy}$ can have different values [Figs.~\ref{fig:mu_dep_spd} (c), (d), (e) and (f)], and $Q^{yy}$ does not show oscillatory behavior at $\theta/\pi=0.99$.
  The details about behaviour of a superfluid stiffness on a square lattice are discussed in Supplemental Material \cite{SM}.

{\it Conclusions.--} We have investigated Meissner effect in non-Hermitian systems of BCS type. For that, we have studied the superfluid stiffness $Q$. In contrast to conventional superconductors, we obtain not just positive real $Q$, but complex in general and with positive and negative signs of $\Re[Q]$ and $\Im[Q]$, depending on the parameters of the system. We argue that the imaginary part comes from the pumping and/or decay of Bogoliubov quasiparticles and the real part is responsible for the type of Meissner effect: conventional diamagnetic or paramagnetic one, the latter being a consequence of instability of superconductivity. An $s$-wave non-Hermitian superconductor can have large negative superfluid stiffness, indicating very strong paramagnetic response, when the mean fields are almost anti-Hermitian, i.e. $\Delta$ is close to $-\bar{\Delta}$. A $p_x$- and $d$-wave non-Hermitian superconductors also exhibit paramagnetic Meissner effect for certain parameters. We believe that these effects must be noticeable in the systems where particles mediating electron-electron interactions are not in equilibrium, e.g., light- \cite{fava:nature24} or acoustic-wave-induced \cite{kornich:scipost22} superconductivity, exciton-mediated superconductivity \cite{plyashechnik:prb23}.

\begin{acknowledgments}
	We acknowledge useful discussions with Fakher Assaad, J\"org Schmalian, and Bj\"orn Trauzettel. This work was supported by the W{\"u}rzburg-Dresden Cluster of Excellence on Complexity and Topology in Quantum Matter (EXC2147, project-id 390858490) and by the DFG (SFB1170). \end{acknowledgments}
	
	\begin{widetext}
\section*{Supplemental Material }

\maketitle
\renewcommand{\theequation}{S\arabic{equation}}
\setcounter{equation}{0}
\renewcommand{\thefigure}{S\arabic{figure}}
\renewcommand{\figurename}{Supplementary Fig.}

\setcounter{figure}{0}
\renewcommand{\thesection}{S\arabic{section}}
\setcounter{section}{0}

In this Supplemental Material, we present additional details and calculations regarding: 1) derivation of electron-electron interaction potential; 2) Hubbard-Stratonovich transformation with non-Hermitian mean fields; 3) the signs of the real and imaginary parts of the superfluid stiffness; 4) the superfluid stiffness on a square lattice; 5) density of states of 2D non-Hermitian superconductors: $s$-, $p_x$-, and $d$-wave superconductors.  

\section{S1. Derivation of electron-electron interaction potential}
In this section, we study, how non-equilibrium bosons mediating electron-electron interactions can lead to complex electron-electron interaction potential. As follows from Eq. (1), there is $S_{\rm int}$ describing interaction of bosons with electrons. Its usual form is
\begin{eqnarray}
\label{eq:Sintbos}
S_{\rm int}(\bar{\psi},\psi,\bar{\varphi},\varphi)=\int d\tau\sum_{{\bf q},{\bf k}} \Gamma_{\bf q}[\varphi_{\bf q}+\bar{\varphi}_{-{\bf q}}]\bar{\psi}_{{\bf k}+{\bf q}}\psi_{\bf k}.
\end{eqnarray} 
Here, for shortness of notation we use 4-vectors for the quantum numbers ${\bf k}=(\omega_n,{\bf p})$, where $\omega_n$ is a Matsubara frequency and ${\bf p}$ is a momentum. If we expand the exponent in the partition function, we obtain:
\begin{eqnarray}
\mathcal{Z}=\int \mathcal{D}[\bar{\psi},\psi,\bar{\varphi},\varphi]e^{-[S_{\rm el}(\bar{\psi},\psi)+S_{\rm int}(\bar{\psi},\psi,\bar{\varphi},\varphi)+S_{\rm bos}(\bar{\varphi},\varphi)]}=\\ \nonumber=\int \mathcal{D}[\bar{\psi},\psi,\bar{\varphi},\varphi]e^{-[S_{\rm el}(\bar{\psi},\psi)+S_{\rm bos}(\bar{\varphi},\varphi)]}[1-S_{\rm int}(\bar{\psi},\psi,\bar{\varphi},\varphi)+\frac{1}{2}S_{\rm int}(\bar{\psi},\psi,\bar{\varphi},\varphi)^2-...].
\end{eqnarray}
Then, we perform the integration over bosonic degrees of freedom, $\bar{\varphi}$, $\varphi$, obtaining:
\begin{eqnarray}
\mathcal{Z}=\int \mathcal{D}[\bar{\psi},\psi]e^{-S_{\rm el}(\bar{\psi},\psi)}[1-\langle S_{\rm int}(\bar{\psi},\psi,\bar{\varphi},\varphi)\rangle+\frac{1}{2}\langle S_{\rm int}(\bar{\psi},\psi,\bar{\varphi},\varphi)^2\rangle-...].
\end{eqnarray}
Usually $S_{\rm bos}$ is quadratic, and then the averages of the odd powers of $S_{\rm int}$ are zero. Once, we reexponentiate the resulting even powers back, we obtain
\begin{eqnarray}
\mathcal{Z}=\int \mathcal{D}[\bar{\psi},\psi]e^{-S_{\rm el}(\bar{\psi},\psi)+\frac{1}{2}\langle S_{\rm int}(\bar{\psi},\psi,\bar{\varphi},\varphi)^2\rangle}.
\end{eqnarray}
Consequently, the effective electron-electron interaction is $S_{\rm e-e}(\bar{\psi},\psi)=-\frac{1}{2}\langle S_{\rm int}(\bar{\psi},\psi,\bar{\varphi},\varphi)^2\rangle$.
Using Eq.~(\ref{eq:Sintbos}), we obtain
\begin{eqnarray}
S_{\rm e-e}=-\frac{1}{2}\int d\tau\sum_{k_1,k_2,q_1,q_2}\Gamma_{{\bf q}_1}\Gamma_{{\bf q}_2}\bar{\psi}_{{\bf k}_1+{\bf q}_1}\psi_{{\bf k}_1}[\langle \bar{\varphi}_{-{\bf q}_1}\varphi_{{\bf q}_2}\rangle+\langle\varphi_{{\bf q}_1}\bar{\varphi}_{-{\bf q}_2}\rangle]\bar{\psi}_{{\bf k}_2+{\bf q}_2}\psi_{{\bf k}_2}=\\ \nonumber=-\frac{1}{\beta L^d}\int d\tau\sum_{k_1,k_2,q_1,q_2}\bar{\psi}_{{\bf k}_1+{\bf q}_1}\psi_{{\bf k}_1} V({\bf q}_1, {\bf q}_2)\bar{\psi}_{{\bf k}_2+{\bf q}_2}\psi_{{\bf k}_2}.
\end{eqnarray}
Often, $\langle \bar{\varphi}_{-{\bf q}_1}\varphi_{{\bf q}_2}\rangle\propto\delta_{-{\bf q}_1,{\bf q}_2}$, $\langle \varphi_{{\bf q}_1}\bar{\varphi}_{-{\bf q}_2}\rangle\propto\delta_{{\bf q}_1,-{\bf q}_2}$ and we obtain simply $V(\bf{q}_1)\delta_{-{\bf q}_1,{\bf q}_2}$. However, the most important is that $\langle \bar{\varphi}_{-{\bf q}_1}\varphi_{{\bf q}_2}\rangle$ is a Green's function. 

If bosons interact with a bath or are under some other external effect that leads to momentum being not a good quantum number, the states $\varphi_{\bf q}$ will decay. This process is included as self-energy into the Green's function, and this will give an imaginary part of it, and consequently an imaginary part of $V({\bf q}_1,{\bf q}_2)$.

Therefore, if bosons mediating electron-electron interaction interact with the bath or experience some other external effect, usually electron-electron interaction $V({\bf q}_1,{\bf q}_2)$ is complex. Then $S_{\rm e-e}$ is non-Hermitian. We note that with this order of Grassmann fields, real but asymmetric $V({\bf q})$ still gives non-Hermitian $S_{\rm e-e}$.

\section{S2. Hubbard-Stratonovich transformation with complex interaction potential}
In this section, we derive Hubbard-Stratonovich transformation in non-Hermitian case. We start from effective electron-electron interaction with Grassmann fields ordered in conventional way:
\begin{eqnarray}
S_{\rm e-e}=-\sum_{\sigma,\sigma'} \int d\tau d\tau' d{\bf r}, d{\bf r}'\bar{\psi}_{\tau,{\bf r},\sigma}\bar{\psi}_{\tau',{\bf r}',\sigma'}V_{\tau-\tau',{\bf r}-{\bf r}'}\psi_{\tau',{\bf r}',\sigma'}\psi_{\tau,{\bf r},\sigma}.
\end{eqnarray}
We express momentum indices in convenient form and anticommute fermionic Grassmann fields so that they resemble Cooper pairing. Thus, the interaction term is:
\begin{eqnarray}
\label{eq:SeeSM}
S_{\rm e-e}=-\frac{1}{\sqrt{\beta L^d}}\sum_{{\bf k}_1,{\bf k}_2,{\bf q},\sigma,\sigma'}\bar{\psi}_{{\bf k}_1,\sigma}\bar{\psi}_{-{\bf k}_1+{\bf q},\sigma'}V_{{\bf k}_1-{\bf k}_2}\psi_{-{\bf k}_2+{\bf q},\sigma'}\psi_{{\bf k}_2,\sigma},
\end{eqnarray}
where indices, e.g. ${\bf k}_1$, are 4-vectors ${\bf k}_1=(\omega_{n_1},{\bf p}_1)$ consisting of a Matsubara frequency $\omega_{n_1}$ and a momentum ${\bf p}_1$. The Fourier transformation is defined as $\psi_{\tau,{\bf r},\sigma}=(L^d\beta)^{-1/2}\sum_{{\bf k},n}\psi_{n,{\bf k},\sigma}e^{i{\bf k}\cdot {\bf r}-i\omega_n \tau}$ and the same for $V_{\tau,{\bf r}}$. 

Then we add a unity (the normalisation is in $\mathcal{D}$) to the partition function $\mathcal{Z}$ (we omit the spin indices for shortness of notation):
\begin{eqnarray}
\label{eq:unity}
&&\mathbb{1}=\int \mathcal{D}[\Delta^\dagger,\Delta]\exp{\left[-\sqrt{\beta L^d}\sum_{{\bf a},{\bf c},{\bf q}}\Delta^{\dagger}_{{\bf q},{\bf a}}(V^\dagger)^{-1}_{{\bf a}-{\bf c}}\Delta_{{\bf c},{\bf q}}\right]}=\\ \nonumber&&=\int \mathcal{D}[\Delta^\dagger,\Delta]\exp{\left[-\sqrt{\beta L^d}\sum_{{\bf a},{\bf b},{\bf b}',{\bf c},{\bf q}}\Delta^\dagger_{{\bf q},{\bf b}'}(V^\dagger)^{-1}_{{\bf b}'-{\bf b}}V_{{\bf b}-{\bf a}}V^{-1}_{{\bf a}-{\bf c}}\Delta_{{\bf c},{\bf q}}\right]}=\int \mathcal{D}[\Delta^\dagger,\Delta]\exp{\left[-\sqrt{\beta L^d}\sum_{{\bf a},{\bf c},{\bf q}}\bar{\Delta}_{{\bf q},{\bf a}}V^{-1}_{{\bf a}-{\bf c}}\Delta_{{\bf c},{\bf q}}\right]}.
\end{eqnarray}
This extended unity exists, i.e. the integral converges, if $V^{-1}$ has a positive definite Hermitian part, that is $[V^{-1}+(V^\dagger)^{-1}]/2$ is positive definite \cite{altland:book10}. Note, that we made a substitution of $\bar{\Delta}$ instead of $\Delta^\dagger$, that is Hermitian conjugate to the bosonic field $\Delta$, under exponent in order to perform the following shift of $\Delta$ and $\bar{\Delta}$:
\begin{eqnarray}
\Delta_{{\bf c},{\bf q}}^{\sigma',\sigma}&\rightarrow& \Delta_{{\bf c},{\bf q}}^{\sigma',\sigma}+\frac{1}{\sqrt{\beta L^d}}\sum_{{\bf b}}V_{{\bf c}-{\bf b}}\psi_{-{\bf b}+{\bf q},\sigma'}\psi_{{\bf b},\sigma},\\
\bar{\Delta}_{{\bf q},{\bf a}}^{\sigma,\sigma'}&\rightarrow&  \bar{\Delta}_{{\bf q},{\bf a}}^{\sigma,\sigma'}+\frac{1}{\sqrt{\beta L^d}}\sum_{{\bf b}'}\bar{\psi}_{{\bf b}',\sigma}\bar{\psi}_{-{\bf b}'+{\bf q},\sigma'}V_{{\bf b}'-{\bf a}}.
\end{eqnarray}
Please, note that $\bar{\Delta}$ and $\Delta$ are in general non-Hermitian. Thus, we obtain under the exponent in Eq.~(\ref{eq:unity}),
\begin{eqnarray}
\sum_{{\bf a},{\bf c},{\bf q},\sigma,\sigma'}\left[-\sqrt{\beta L^d}\bar{\Delta}_{{\bf q},{\bf a}}^{\sigma,\sigma'}V^{-1}_{{\bf a}-{\bf c}}\Delta_{{\bf c},{\bf q}}^{\sigma',\sigma}-\bar{\psi}_{{\bf c},\sigma}\bar{\psi}_{-{\bf c}+{\bf q},\sigma'}\Delta_{{\bf c},{\bf q}}^{\sigma',\sigma}-\bar{\Delta}_{{\bf q},{\bf a}}^{\sigma,\sigma'}\psi_{-{\bf a}+{\bf q},\sigma'}\psi_{{\bf a},\sigma}-\right.\\ \nonumber\left.-\frac{1}{\sqrt{\beta L^d}}\bar{\psi}_{{\bf c},\sigma}\bar{\psi}_{-{\bf c}+{\bf q},\sigma'}V_{{\bf c}-{\bf a}}\psi_{-{\bf a}+{\bf q},\sigma'}\psi_{{\bf a},\sigma}\right],
\end{eqnarray}
 where the last term cancels exactly with $S_{\rm e-e}$ from Eq.~(\ref{eq:SeeSM}). Later, one can simplify the model to consider only zero center-of-mass momentum Cooper pairs.  
 
\section{S3. The signs of $\Re{[Q(\mu=0)]}$ and $\Im{[Q(\mu=0)]}$}
Here, we present the calculation about the signs of $\Re{[Q(\mu=0)]}$ and $\Im{[Q(\mu=0)]}$. Eq.~(10) of the main text reads:
\begin{eqnarray}
Q(\mu=0)=\frac{ie^2\sqrt{2m}\bar{\Delta}\Delta}{6\pi^2}{\rm sign}^*[\Im[\bar{\Delta}\Delta]] \int d\omega \left[\frac{(-\bar{\Delta}\Delta-\omega^2)^{1/4}-\sqrt{-\sqrt{-\bar{\Delta}\Delta-\omega^2}}}{\bar{\Delta}\Delta+\omega^2}\right].
\end{eqnarray}
If we expand the square roots in this expression and take its principal values, we obtain
\begin{eqnarray}
\label{eq:Qmu0}
Q(\mu=0)=\frac{e^2\sqrt{m}}{6\pi^2} \int d\omega\frac{\bar{\Delta}\Delta}{\bar{\Delta}\Delta+\omega^2}\times \\ \nonumber\left[\sqrt{[(\Re[\bar{\Delta}\Delta]+\omega^2)^2+\Im[\bar{\Delta}\Delta]^2]^{1/4}+\sqrt{\frac{1}{2}(\sqrt{(\Re[\bar{\Delta}\Delta]+\omega^2)^2+\Im[\bar{\Delta}\Delta]^2}-\Re[\bar{\Delta}\Delta]-\omega^2)}}(1+i{\rm sign}^*[\Im(\bar{\Delta}\Delta)])+\right.\\ \nonumber\left.+(1-i{\rm sign}^*[\Im(\bar{\Delta}\Delta)])\sqrt{[(\Re[\bar{\Delta}\Delta]+\omega^2)^2+\Im[\bar{\Delta}\Delta]^2]^{1/4}-\sqrt{\frac{1}{2}(\sqrt{(\Re[\bar{\Delta}\Delta]+\omega^2)^2+\Im[\bar{\Delta}\Delta]^2}-\Re[\bar{\Delta}\Delta]-\omega^2)}}\right].
\end{eqnarray}
We denote the terms in the square brackets as $A+i{\rm sign}^*[\Im{(\bar{\Delta}\Delta)}]B$, where $A,B\geq 0$. Taking into account that the integral over $\omega$ contains only $\omega^2$ we can reduce it to $2\int_0^\infty ... d\omega$. Thus, if the real (or imaginary) part of function under integral in Eq.~(\ref{eq:Qmu0}) is positive then $\Re{[Q(\mu=0)]}$ (or $\Im{[Q(\mu=0)]}$) is also positive. The same holds for negative values. 

The real part of the function under integral in Eq.~(\ref{eq:Qmu0}) is
\begin{eqnarray}
\frac{|\bar{\Delta}\Delta|^2A+(\Re[\bar{\Delta}\Delta]A-|\Im[\bar{\Delta}\Delta]|B)\omega^2}{(\Re[\bar{\Delta}\Delta]+\omega^2)^2+\Im[\bar{\Delta}\Delta]^2}.
\end{eqnarray}
Thus, when $\Re[\bar{\Delta}\Delta]A>|\Im[\bar{\Delta}\Delta]|B$, the real part is definitely positive. If not, it becomes an interplay with the term $|\bar{\Delta}\Delta|^2A$. In order for $\Re[\bar{\Delta}\Delta]A>|\Im[\bar{\Delta}\Delta]|B$ to hold, we need first of all $\Re[\bar{\Delta}\Delta]>0$. Then, we can see from Eq.~(\ref{eq:Qmu0}), that $A>B$. Thus, $\Re[\bar{\Delta}\Delta]>0$ guarantees that $\Re{[Q(\mu=0)]}>0$ and most likely then $\Re[Q]>0$, because usually $\Re{[Q]}$ grows with $\mu$.

If $\Re[Q(\mu=0)]<0$, then the second large square root in the square brackets in Eq. (\ref{eq:Qmu0}) can become imaginary for certain $\omega$. We have $A$ and $B$ slightly restructured. Unfortunately, due to the complexity of interplay with $A|\bar{\Delta}\Delta|^2$, we cannot derive analytically, when the transition from $\Re[Q(\mu=0)]>0$ to $\Re[Q(\mu=0)]<0$ occurs.

Now, let's consider imaginary part of the function under integral in Eq.~(\ref{eq:Qmu0}). It is,
\begin{eqnarray}
\frac{i{\rm sign}^*[\Im(\bar{\Delta}\Delta)][(|\Im[\bar{\Delta}\Delta]|A+\Re[\bar{\Delta}\Delta]B)\omega^2+|\bar{\Delta}\Delta|^2B]}{(\Re[\bar{\Delta}\Delta]+\omega^2)^2+\Im[\bar{\Delta}\Delta]^2}.
\end{eqnarray}
If $\Re[\bar{\Delta}\Delta]>0$, then the sign of it depends solely on $\Im[\bar{\Delta}\Delta]$. If $\Re[\bar{\Delta}\Delta]<0$, then we again obtain complicated situation with an interplay between terms, which we cannot estimate analytically.

\section{S4. Superfluid Stiffness on a square lattice}
As written in the main text, we consider $s$-, $p_x$-, and $d$-wave superconductors on the square lattice, where the Hamiltonian is given by
\begin{align}
    H_\mathbf{k}(\mathbf{A})
    =&
    \begin{pmatrix}
      \varepsilon_{\mathbf{k}-e\mathbf{A}} &\Delta(\mathbf{k})
      \\
      \Delta(\mathbf{k}) e^{i\theta} & -\varepsilon_{\mathbf{k}+e\mathbf{A}}
    \end{pmatrix}
\end{align}
with $0\leq \theta<\pi$ and
\begin{align}
  \varepsilon_{\mathbf{k}\pm e\mathbf{A}}
  =&
  -2t[\cos (k_x\pm eA_x)+\cos (k_y\pm eA_y)]-\mu,
  \\
  \Delta(\mathbf{k})
  =&
  \begin{cases}
    \Delta_0 &s\text{-wave},\\
    \Delta_0\sin k_x &p_x\text{-wave},\\
    \Delta_0(\cos k_x-\cos k_y) &d\text{-wave}.
  \end{cases}
\end{align}
The eigenvalues are given by
\begin{align}
  E_{\mathbf{k},\pm}(\mathbf{A})
  =&
  \pm\sqrt{{\{2t[\cos k_x\cos (eA_x)+\cos k_y\cos(eA_y)]+\mu\}}^2 + \Delta^2(\mathbf{k})e^{i\theta}}
  +2t\sin k_x\sin(eA_x)+2t\sin k_y\sin(eA_y).
  \label{eq:eigenvalues_sm}
\end{align}
The action is given by 
\begin{align}
  S(\mathbf{A})=-\sum_{n,\mathbf{k},\alpha=\pm}\mathrm{Tr}\ln [E_{\mathbf{k},\alpha}(\mathbf{A})-i\omega_n]e^{i\omega_n\eta}
  \label{eq:action_sm}
\end{align}
with a positive infinitesimal $\eta$ and $\omega_n=(2n+1)\pi/\beta$ ($n\in\mathbb{Z}$).
In Eq.~\eqref{eq:action_sm}, we suppose $E_{\mathbf{k},\pm}(\mathbf{A})\neq i\omega_n$.
$E_{\mathbf{k},\pm}(\mathbf{A})$ is real or purely imaginary when $\theta=\pi$.
Hence, there are two cases at $\theta=\pi$:
  \begin{itemize}
    \item 
      When $\max\Im[E_{\mathbf{k},\pm}(\mathbf{A})]<\pi/\beta$, there is no singularity in $S(\mathbf{A})$, and $Q^{ab}$ is well defined.
    \item 
      When $\max\Im[E_{\mathbf{k},\pm}(\mathbf{A})]\geq \pi/\beta$ and the purely imaginary eigenvalues are continuous from zero to $\max\Im[E_{\mathbf{k},\pm}(\mathbf{A})]$, $E_{\mathbf{k},\pm}(\mathbf{A})=i\omega_n$ is realized at some $\mathbf{k}$.
      Hence, the action, as well as $Q^{ab}$, is ill-defined.
      Continuous eigenvalue corresponds to $L\rightarrow\infty$.
      On the other hand, if $L$ is finite, $\max\Im[E_{\mathbf{k},\pm}(\mathbf{A})]\geq \pi/\beta$ is a necessary condition to realize $E_{\mathbf{k},\pm}(\mathbf{A})=i\omega_n$.
      Therefore, for finite $L$, the action can be well defined when $E_{\mathbf{k},\pm}(\mathbf{A})\neq i\omega_n$.
  \end{itemize}
  The action can be expanded with respect to $\mathbf{A}$ as
  \begin{align}
    S(\mathbf{A})
    =&
    S(\mathbf{A}=0)
    +
    \sum_{a=x,y}
    \frac{\partial S(\mathbf{A})}{\partial A_a}\Big|_{\mathbf{A}=0} A_a 
    +
    \frac{1}{2}
    \sum_{a,b=x,y}
    \frac{\partial^2 S(\mathbf{A})}{\partial A_a\partial A_b}\Big|_{\mathbf{A}=0}
    A_a A_b
    + \mathcal{O}(\mathbf{A}^3).
  \end{align}
  Here, the first and the second derivative of the action are given by
  \begin{align}
    \frac{\partial S(\mathbf{A})}{\partial A_a}\Big|_{\mathbf{A}=0} 
    =&
    -\sum_{n,\mathbf{k},\alpha=\pm}\frac{e^{i\omega_n\eta}}{E_{\mathbf{k},\alpha}(\mathbf{A})-i\omega_n}
    \frac{\partial E_{\mathbf{k},\alpha}(\mathbf{A})}{\partial A_a}\Big|_{\mathbf{A}=0}
    \nonumber\\
    =&
    \sum_{\mathbf{k},\alpha=\pm}\frac{\beta}{1+e^{\beta E_{\mathbf{k},\alpha}(\mathbf{A})}}
    \frac{\partial E_{\mathbf{k},\alpha}(\mathbf{A})}{\partial A_a}\Big|_{\mathbf{A}=0},
    \\
    \frac{\partial^2 S(\mathbf{A})}{\partial A_a\partial A_b}\Big|_{\mathbf{A}=0} 
    =&
    \sum_{\mathbf{k},\alpha=\pm}
    \frac{-\beta^2}{4\cosh^2\frac{\beta E_{\mathbf{k},\alpha}(\mathbf{A})}{2}}
    \frac{\partial E_{\mathbf{k},\alpha}(\mathbf{A})}{\partial A_a}
    \frac{\partial E_{\mathbf{k},\alpha}(\mathbf{A})}{\partial A_b}
    \Big|_{\mathbf{A}=0}
    +
    \sum_{\mathbf{k},\alpha=\pm}
    \frac{\beta}{1+e^{\beta E_{\mathbf{k},\alpha}(\mathbf{A})}}\frac{\partial^2 E_{\mathbf{k},\alpha}(\mathbf{A})}{\partial A_a \partial A_b}
    \Big|_{\mathbf{A}=0},
  \end{align}
  respectively.
  The derivatives of eigenvalues are given by
  \begin{align}
    \frac{\partial E_{\mathbf{k},\pm}(\mathbf{A})}{\partial A_x}\Big|_{\mathbf{A}=0}
    =&
    -
    2te \sin k_x,
    \label{eq:dedax}
    \\
    \frac{\partial E_{\mathbf{k},\pm}(\mathbf{A})}{\partial A_y}\Big|_{\mathbf{A}=0}
    =&
    -
    2te \sin k_y,
    \label{eq:deday}
    \\
    \frac{\partial^2 E_{\mathbf{k},\pm}(\mathbf{A})}{\partial A_x^2}\Big|_{\mathbf{A}=0}
    =&
    \frac{2te^2\cos k_x [-2t(\cos k_x + \cos k_y)-\mu]}{E_{\mathbf{k},\pm}(\mathbf{A}=0)},
    \\
    \frac{\partial^2 E_{\mathbf{k},\pm}(\mathbf{A})}{\partial A_y^2}\Big|_{\mathbf{A}=0}
    =&
    \frac{2te^2\cos k_y [-2t(\cos k_x + \cos k_y)-\mu]}{E_{\mathbf{k},\pm}(\mathbf{A}=0)},
    \\
    \frac{\partial^2 E_{\mathbf{k},\pm}(\mathbf{A})}{\partial A_x\partial A_y}\Big|_{\mathbf{A}=0}
    =&
    0.
    \label{eq:dEdxdy}
  \end{align}
  From Eq.~\eqref{eq:eigenvalues_sm}, $E_{\mathbf{k},\pm}(\mathbf{A}=0)$ is an even function of $k_x$ and $k_y$, and from Eqs.~\eqref{eq:dedax} and \eqref{eq:deday}, $\partial E_{\mathbf{k},\pm}(\mathbf{A})/\partial A_{x(y)}|_{\mathbf{A}=0}$ is an odd function of $k_x(k_y)$.
  Hence, the first derivative of the action is zero, and the expansion of the action starts from the second derivative of the action.
  The superfluid stiffness is given by
  \begin{align}
    Q^{ab}
    =&
    \frac{1}{\beta L^2}
    \frac{\partial^2 S(\mathbf{A})}{\partial A_a\partial A_b}\Big|_{\mathbf{A}=0}
    \nonumber\\
    =&
    Q^{(1)ab} + 
    Q^{(2)ab},
    \\
    Q^{(1)ab}
    =&
    \frac{1}{L^2}
    \sum_{\mathbf{k},\alpha=\pm}
    \frac{-\beta}{4\cosh^2\frac{\beta E_{\mathbf{k},\alpha}(\mathbf{A})}{2}}
    \frac{\partial E_{\mathbf{k},\alpha}(\mathbf{A})}{\partial A_a}
    \frac{\partial E_{\mathbf{k},\alpha}(\mathbf{A})}{\partial A_b}
    \Big|_{\mathbf{A}=0},
    \\
    Q^{(2)ab}
    =&
    \frac{1}{L^2}
    \sum_{\mathbf{k},\alpha=\pm}
    \frac{1}{1+e^{\beta E_{\mathbf{k},\alpha}(\mathbf{A})}}
    \frac{\partial^2 E_{\mathbf{k},\alpha}(\mathbf{A})}{\partial A_a \partial A_b}
    \Big|_{\mathbf{A}=0}.
  \end{align}
  $Q^{xy}$ is zero since $\cosh\frac{\beta E_{\mathbf{k},\pm}(0)}{2}$ is an even function of $k_x$ and $k_y$, and $\frac{\partial E_{\mathbf{k},\pm}(\mathbf{A})}{\partial A_x}\frac{\partial E_{\mathbf{k},\pm}(\mathbf{A})}{\partial A_y}\big|_{\mathbf{A}=0}$ is an odd function of $k_x$ and $k_y$.

  At $\theta=\pi$, if $E_{\mathbf{k},\pm}(\mathbf{A})\neq i\omega_n$ for any $\mathbf{k}$, $Q^{(1)xx}$ and $Q^{(1)yy}$ are real and non-positive.
  All eigenvalues are real or purely imaginary at $\theta=\pi$, and $\cosh^2\frac{\beta E_{\mathbf{k},\pm}}{2}$ is real and non-negative.
  ${\left(\frac{\partial E_{\mathbf{k},\pm}(\mathbf{A})}{\partial A_a}\right)}^2\big|_{\mathbf{A}=0}$ is also real and non-negative.
  Hence, $Q^{(1)xx}$ and $Q^{(1)yy}$ are non-positive values.
  If some eigenvalues are close to $i\omega_n$, the denominators of $Q^{(1)ab}$ goes to zero.
  In this case, the denominator of $Q^{(2)ab}$ also approaches zero, but the sign of $Q^{(2)ab}$ depends on $\mathbf{k}$ and other parameters.

\subsection{A. $s$-wave case}
\begin{figure}[t]
  \begin{center}
    \includegraphics[width=7.5cm]{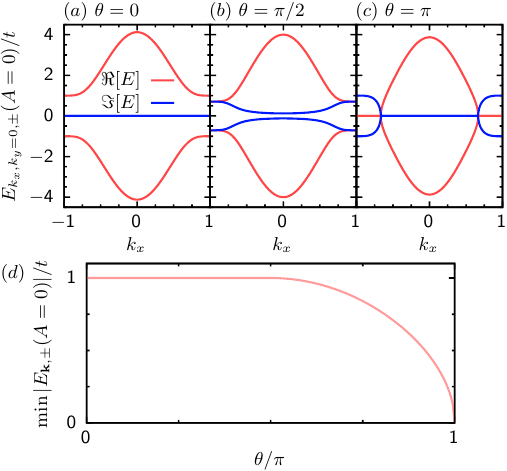}
  \caption{
    The energy dispersion for (a) $\theta=0$, (b) $\pi/2$, and (c) $\pi$ is plotted as a function of $k_x$ at $k_y=0$.
    (d) $\min|E_{\mathbf{k},\pm}(0)|$ is plotted as a function of $\theta$.
    $\Delta_0/t=1$ and $\mu=0$ for all figures.
  }\label{fig:dispersion}
  \end{center}
\end{figure}
Let us take a close look at the $s$-wave case.
In this case, the energy gap can only close at $\theta=\pi$ as shown in Fig.~\ref{fig:dispersion}.
For $\theta/\pi\lesssim0.5$, the energy gap is almost constant, and for $\theta/\pi\gtrsim0.5$, the energy gap decreases as $\theta$ increases [Fig.~\ref{fig:dispersion} (d)].
\begin{figure}[b]
  \begin{center}
    \includegraphics[width=12.9cm]{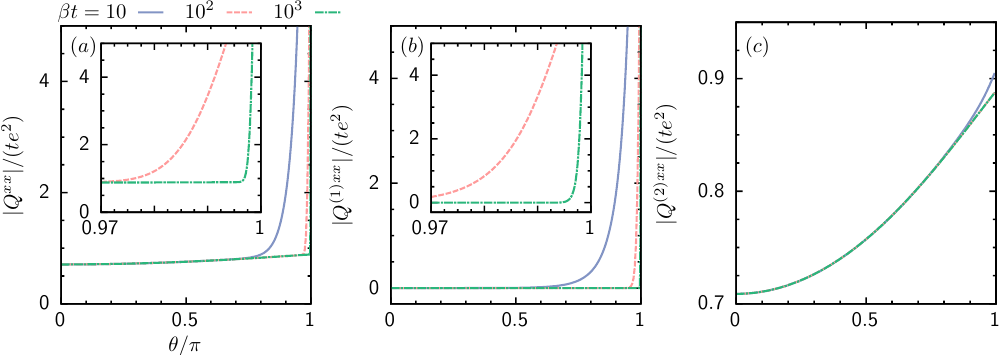}
  \caption{
    Absolute value of (a) $Q^{xx}$, (b) $Q^{(1)xx}$, and (c) $Q^{(2)xx}$ are plotted as a function of $\theta$ for $s$-wave superconductor with several values of $\beta$.
    The insets in (a) and (b) shows magnified views close to $\theta=\pi$.
    $\Delta_0/t=1$ and $\mu=0$ for all figures.
  }\label{fig:Q_decompose}
  \end{center}
\end{figure}
\begin{figure}[th]
  \begin{center}
    \includegraphics[width=12.9cm]{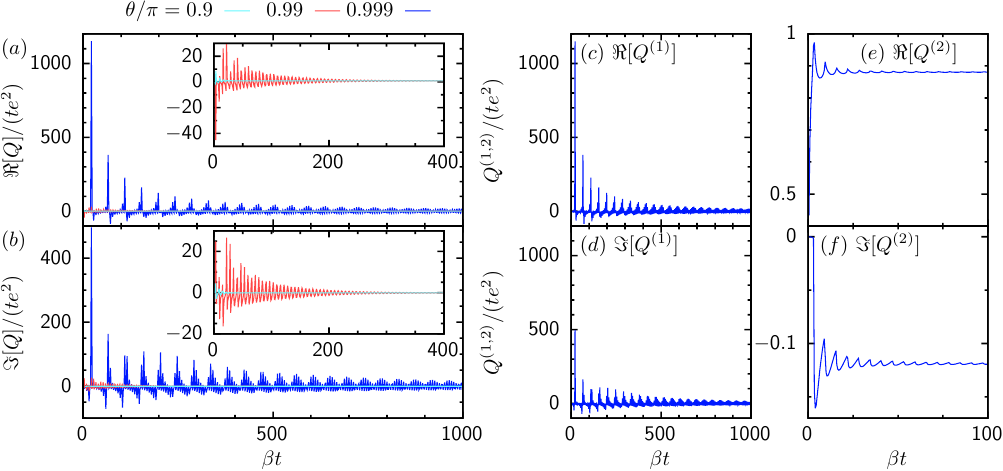}
  \caption{
    Temperature dependence of $Q$ is shown for $\theta/\pi=0.9$, $0.99$, and $0.999$.
    (a) Real part and (b) imaginary parts are shown.
    In the insets of (a) and (b), magnified view of them are plotted.
    (c) $\Re[Q^{(1)}]$, 
    (d) $\Im[Q^{(1)}]$, 
    (e) $\Re[Q^{(2)}]$, and
    (f) $\Im[Q^{(2)}]$, 
    are plotted as a function of $\beta$ for $\theta/\pi=0.999$.
    $\Delta_0/t=1$ and $\mu=0$ for all plots.
  }\label{fig:beta_dep}
  \end{center}
\end{figure}
We show the temperature dependence of $Q^{xx}$ in Figs.~\ref{fig:Q_decompose} and \ref{fig:beta_dep}.
The temperature dependence of $Q^{(1)xx}$ is conspicuous compared with $Q^{(2)xx}$, i.e., $Q^{(2)}$ for $\beta t=10^2$ and $10^3$ are almost identical, and $Q^{(2)}$ at $\beta t=10$ deviates from that at $\beta t=10^2$ and $10^3$ for $\theta/\pi\gtrsim0.8$ [Fig.~\ref{fig:Q_decompose}].
$\beta$ dependence of $Q$ is shown in Fig.~\ref{fig:beta_dep} close to $\theta=\pi$.
We can see oscillation in Figs.~\ref{fig:beta_dep} (a) and (b), and the amplitude of this oscillation becomes smaller as $\beta$ becomes larger.
As $\theta$ approaches $\pi$, the dumping of the oscillation becomes slower.
$Q^{(1)xx}$ and $Q^{(2)xx}$ are shown in Figs.~\ref{fig:beta_dep} (c), (d), (e), and (f).

\section{S5. Density of states for 2D non-Hermitian superconductors}

In this section, we will consider the density of states of 2D $s$-, $p_x$-, and $d$-wave superconductors. The spectrum of an $s$-wave superconductor at different $\theta$ was presented in Figs.~\ref{fig:dispersion} (a)-(c). Here, we present the spectra of $p_x$- and $d$-wave superconductors at different $\theta$, Figs.~\ref{fig:pspectrum}, \ref{fig:dspectrum}. \begin{figure}[h]
\begin{center}
    \includegraphics[width=8cm]{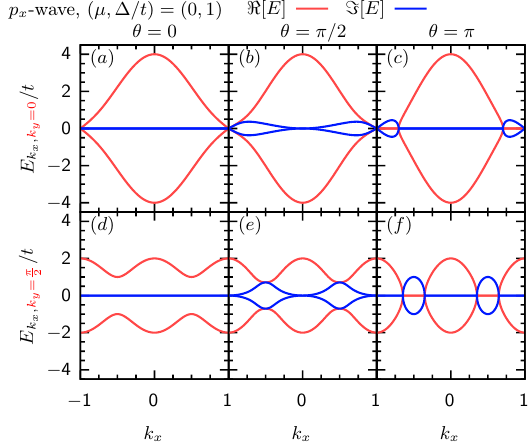}
  \caption{
    The energy spectrum of a 2D $p_x$-wave superconductor for (a), (d) $\theta=0$, (b), (e) $\theta=\pi/2$, and (c), (f) $\theta=\pi$ is plotted as a function of $k_x$ at (a), (b), (c) $k_y=0$ and (d), (e), (f) $k_y=\pi/2$.
  }\label{fig:pspectrum}
  \end{center}
\end{figure}
\begin{figure}[h]
\begin{center}
    \includegraphics[width=8cm]{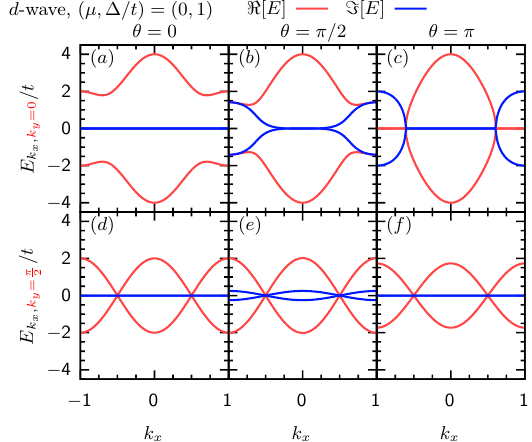}
  \caption{
     The energy spectrum of a 2D $d$-wave superconductor for (a), (d) $\theta=0$, (b), (e) $\theta=\pi/2$, and (c), (f) $\theta=\pi$ is plotted as a function of $k_x$ at (a), (b), (c) $k_y=0$ and (d), (e), (f) $k_y=\pi/2$.  }\label{fig:dspectrum}
  \end{center}
\end{figure}

The density of states (DOS) for non-Hermitian superconductors are given by
\begin{align}
  D(E)=\frac{1}{N}\sum_\mathbf{k}\delta(\Re[E]-\Re[E_\mathbf{k}])\delta(\Im[E]-\Im[E_\mathbf{k}]).
\end{align}
Here, $E_\mathbf{k}=\pm\sqrt{\varepsilon_\mathbf{k}^2+\Delta(k)^2e^{i\theta}}$ is the eigenvalue of the Hamiltonian with $\varepsilon_\mathbf{k}=-2t(\cos k_x+\cos k_y)-\mu$ for 2D SCs.

\begin{figure}[h]
\begin{center}
    \includegraphics[width=15cm]{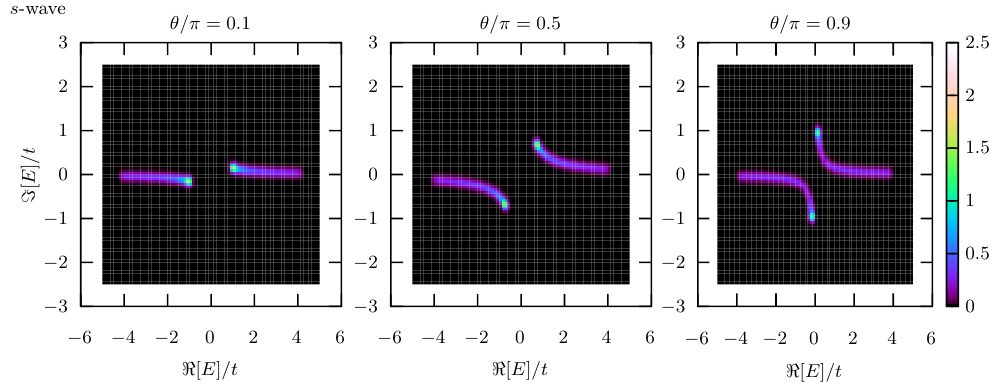}
  \caption{
    The DOS for $s$-wave SC is shown as function  of $\Re[E]$ and $\Im[E]$ with $\Delta_0/t=1$ and $\mu=0$ for $\theta/\pi=0.1$, $0.5$, and $0.9$.
  }\label{fig:dos_s}
  \end{center}
\end{figure}
\begin{figure}[h]
  \begin{center}
    \includegraphics[width=15cm]{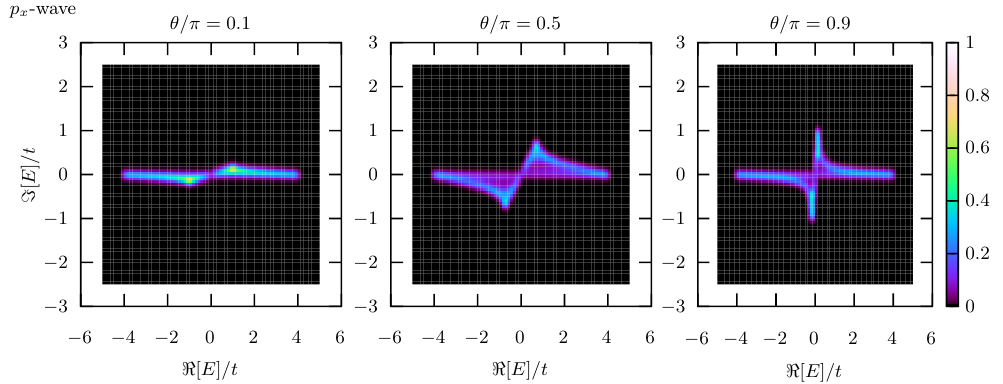}
  \caption{
    The DOS for $p_x$-wave SC is shown as function  of $\Re[E]$ and $\Im[E]$ with $\Delta_0/t=1$ and $\mu=0$ for $\theta/\pi=0.1$, $0.5$, and $0.9$.
  }\label{fig:dos_p}
  \end{center}
\end{figure}
\begin{figure}[h]
  \begin{center}
    \includegraphics[width=15cm]{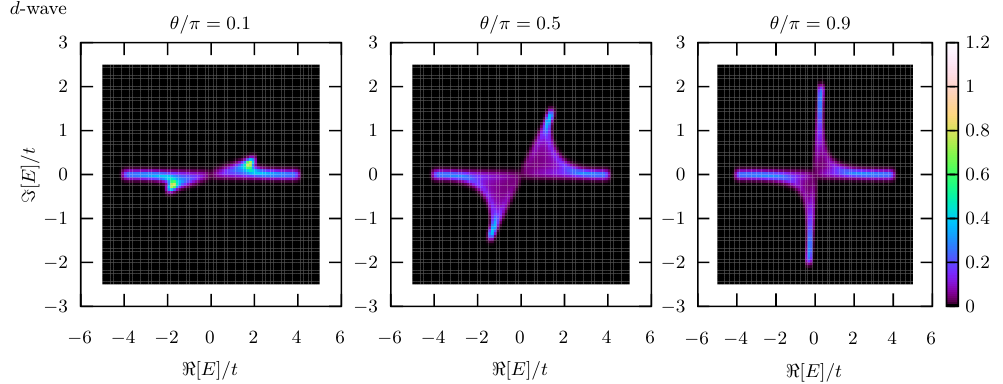}
  \caption{
    The DOS for $d$-wave SC is shown as function  of $\Re[E]$ and $\Im[E]$ with $\Delta_0/t=1$ and $\mu=0$ for $\theta/\pi=0.1$, $0.5$, and $0.9$.
  }\label{fig:dos_d}
  \end{center}
\end{figure}

We show the DOS for 2D $s$-, $p_x$-, and $d$-wave in Figs.~\ref{fig:dos_s}, \ref{fig:dos_p}, and \ref{fig:dos_d}, respectively.
As $\theta$ increases, the DOS deviates from the real axis. We can see that for all of them, $s$-, $p_x$- and $d$-wave superconductors, the imaginary part of the DOS is close to zero for $\theta=0.1$, that reflects zero imaginary part of energy in their spectra at $\theta=0$. Analogously, at $\theta=0.9$, their DOS have almost orthogonal imaginary and real parts corresponding to orthogonal imaginary and real parts of energy at $\theta=\pi$.

The DOS of all considered superconductors exhibit peaks similar to Hermitian superconductors of these types.

\ \
\ \

\ \

\ \

\ \

 \ \


\end{widetext}

\end{document}